\documentclass[10 pt, conference]{IEEEtran}
\pdfoutput=1

\usepackage[utf8]{inputenc}
\usepackage[T1]{fontenc}
\usepackage{graphicx}
\usepackage{breqn}
\usepackage{pgfplots}
\usepackage{subcaption}
\usepackage{amsmath}
\usepackage{amssymb}
\usepackage{amsthm}
\usepackage{url}
\usepackage{footnote}

\usepackage[ruled,vlined]{algorithm2e}

\usepackage[normalem]{ulem}

\numberwithin{theorem}{subsection}

\usepackage{mathtools}

\usepackage{stackengine}
\newcommand\underparen[1]{\@ifnextchar_{\uphelp{\uparen{#1}}}{\uparen{#1}}}
\makeatother
\def\uphelp#1_#2{\ensurestackMath{\stackunder[1pt]{#1}{\scriptstyle #2}}}
\newcommand\uparen[1]{\setbox0=\hbox{$#1$}\ensurestackMath{%
  \stackunder[0pt]{#1}{\rotatebox{90}{$\left(%
  \rule[\dimexpr-.5\wd0+\dp\strutbox-1.3pt]{0pt}{\wd0}\right.$}}%
}}

\usepackage{amssymb} 
\newcommand{\defeq}[0]{\triangleq}
\usepackage{xspace}

\newcommand{\keepcomment}{1} 

\usepackage[normalem]{ulem}

\ifnum\keepcomment=1
	\usepackage[colorinlistoftodos,textsize=scriptsize]{todonotes} 
    \newcommand{\stkout}[1]{\ifmmode\text{\sout{\ensuremath{#1}}}\else\sout{#1}\fi}
    
\else
	
	\usepackage[disable]{todonotes} 
\fi

\usepackage{paralist} 

\title{
Cache Allocation in Multi-Tenant Edge Computing via online Reinforcement Learning
}

\author{
    \IEEEauthorblockN{Ayoub Ben-Ameur, Andrea Araldo, Tijani Chahed}
    \IEEEauthorblockA{Institut Polytechnique de Paris; Télécom SudParis\\
    \{first\_name\}.\{last\_name\}@telecom-sudparis.com}
}

\usepackage{array}
\newcolumntype{C}[1]{>{\centering\arraybackslash}m{#1}}
\usepackage{pgfplots}
\pgfplotsset{width=8cm,compat=1.9}

\begin{document}

\maketitle
\thispagestyle{plain}
\pagestyle{plain}

\begin{abstract}
  We consider in this work Edge Computing (EC) in a multi-tenant environment: the resource owner, i.e., the Network Operator (NO), virtualizes the resources and lets third party Service Providers (SPs - tenants) run their services, which can be diverse and with heterogeneous requirements. Due to confidentiality guarantees, the NO cannot observe the nature of the traffic of SPs, which is encrypted. This makes resource allocation decisions challenging, since they must be taken based solely on observed monitoring information.
  
  We focus on one specific resource, i.e., cache space, deployed in some edge node, e.g., a base station. We study the decision of the NO about how to partition cache among several SPs in order to minimize the upstream traffic. Our goal is to optimize cache allocation using purely data-driven, model-free Reinforcement Learning (RL). Differently from most applications of RL, in which the decision policy is learned offline on a simulator, we assume no previous knowledge is available to build such a simulator. We thus apply RL in an \emph{online} fashion, i.e., the policy is learned by directly perturbing the actual system and monitoring how its performance changes. Since perturbations generate spurious traffic, we also limit them. We show in simulation that our method rapidly converges toward the theoretical optimum, we study its fairness, its sensitivity to several scenario characteristics and compare it with a method from the state-of-the-art.
  Our code to reproduce the results is available as open source.\footnote{https://github.com/Ressource-Allocation/Cache-Allocation-Project}
\end{abstract}

\maketitle

\section{Introduction}
Data generation rate is expected to exceed the capacity of today’s Internet in the near future~\cite{Wang2019}. It thus becomes more and more important to serve user requests, whenever possible, directly at the edge of the network, thus reducing the upstream traffic, i.e., traffic to/from remote server locations, as well as latency. Hence, Edge Computing (EC) consists in deploying computational capabilities, e.g. RAM, CPU, storage, GPUs, into nodes at the network's edge. Such nodes could be co-located with (micro) base stations, access points, etc.

We position our work in the framework of multi-tenant EC~\cite{AraldoSAC2020, Spallina2022}: the NO owns computational resources at the edge and virtualizes, partitions and allocates them to third party Service Providers (SPs), e.g. YouTube, Netflix, etc. Each SP can then use its assigned share as if it were a dedicated hardware.

We focus in this paper on one resource in particular, namely storage, which is remarkably relevant: indeed, more than 80\% of the Internet traffic might be represented by content delivery, and in particular video~\cite{Cisco}. We assume that the NO owns storage at the edge nodes and uses it as cache. However, the NO cannot operate caching directly, as classically assumed, since all the traffic is encrypted by the SPs. Therefore, it is neither possible for the NO to know which objects are requested, for instance which ones are the most popular, nor whether they are even cacheable, for instance online video broadcast. We thus assume that the NO allocates storage among SPs and lets each SP decide what to cache within the allocated space, as depicted in Fig.~\ref{fig:upstream-traffic}.\footnote{As in classic content caching, we assume the SPs do not pay the NO for the cache: cache is used by SPs for free, and the NO compensates the initial storage deployment cost with upstream traffic reduction.}
Our aim is to solve the problem for the NO to optimally decide how many \emph{cache slots} should be allocated to each SP, in order to minimize the upstream traffic, i.e., the traffic from the Internet to the edge node.
Due to the encrypted nature of traffic, the NO can only base its decision on data-driven strategies consisting in \emph{trial and error}: the NO continuously perturbs the cache allocation and observes the induced variation on the upstream traffic.

We propose a data-driven approach based on RL, used in an \emph{online fashion}: while usually RL is trained offline and then applied to a real system, we instead train RL directly on the system while it is up and running. Therefore, we are not only interested in finding a good cache allocation, but also in \emph{how} to find it. Indeed, while the only way for the NO to learn how to optimize the allocation is to continuously perturb it, we also need to keep the cost of such perturbations reasonable. 
We conduct simulation-based experiments and show that our RL optimization approaches the theoretical optimal allocation and outperforms a state-of-the art method.

The remainder of this paper is organized as follows. In section~\ref{sec:related-work}, we review some works related to our present topic. In section~\ref{sec:system-model}, we present our system and model. In section~\ref{sec:rl-formulation}, we formulate our problem using RL. In section~\ref{sec:performance-evaluation}, we show our simulation results. Section~\ref{sec:concl} contains  our conclusion and some hints on future works.  
\section{Related Work}
\label{sec:related-work}

Authors of~\cite{Towsley2018} show that a utility driven cache partitioning outperforms sharing it. However, they need information about the system conditions in order to solve it. We assume instead that no information is available and that optimization is done by observing the changes in upstream traffic induced by perturbing the allocation. In~\cite{Secci2020a}, authors consider the difference between the resources demanded by each SP and the resources actually allocated with the aim to be fair. In \cite{sahar2016}, the authors propose a resource pricing framework for one NO and several SPs, for several well-established resource allocations knowing the demand of users. We instead do not know anything about the requests nature. Also, our focus is on resource allocation and not pricing; SPs do not pay for the resources.
In~\cite{Gao2020}, the mobile edge network is assumed to have multiple cache servers to assist SPs, each with its own set of users (while we do not limit a user to a single SP) and acts as a rational selfish player, in a bargaining game, aiming to maximize its utility. ~\cite{Ahmadi2019} also considers sharing cache between SPs, by applying coalitional game theory. Different from them, our allocation decision is centralized by the NO and we do not require any payment.

RL has been used for resource allocation in the context of EC in, for instance, \cite{Wang2019b},~\cite{Rao2009},~\cite{Kandula2016} and~\cite{Zhou2017}. Contrary to our approach, authors pre-train the RL algorithm offline on a simulated system before using it on the running system. We instead do not have information to build a simulator, we train our algorithm online. This imposes on us a more parsimonious learning strategy.
In \cite{jiang2020}, authors present a RL algorithm for resource auto-scaling in clouds: resources are assumed to be unlimited, however the goal is to allocate to each SP an amount of resources that does not exceed its needs. In our case instead, resources are scarce, allocating resources to one SP means allocating less for another. 
To the best of our knowledge, the only method we can compare against is Simultaneous Perturbation Stochastic Approximation (SPSA)~\cite{Araldo2018}, since the latter is the only work to propose a data-driven approach to partition cache among several SPs. They do so based on stochastic optimization. However, they need to continuously perturb the allocation, generating spurious upstream traffic that may be non-negligible. We instead include traffic perturbation into the optimization objective, thus managing to keep it low, which allows us to outperform~\cite{Araldo2018}, as shown in section~\ref{sec:performance-evaluation}.

\section{System Model}
\label{sec:system-model}

We consider a setting with one NO, owning the resources, cache in our case, and willing to share them between $P$ SPs. 

\subsection{Request Pattern}
\label{sec:Request-patterm}
Requests of users arrive with rate $\lambda$ expressed in $req/s$. Each request is directed to one of the $P$ SPs. Let $f_p$ denote the probability that a given request is for SP $p$. Therefore, $\lambda\cdot f_p$ is the request arrival rate for SP $p$.
The objects of the SPs are not all eligible to be stored in the cache (e.g., live streams and boadcasts). To represent this, each SP $p$ has a certain \emph{cacheability} $\zeta_p$, which is the probability that the user request is for a cacheable content. 
We consider that SP $p$ has a catalog of $N_p$ cacheable objects. We denote each object as $(c,p)$, where $c=1,2,\dots,N_p$ is the identifier of the object within SP $p$. Each object of SP $p$ has its own popularity $\rho_{c,p}$ which is defined as the probability that, taking any request for a cacheable object of SP $p$, that object is $c$. Therefore, each cacheable object $c$ of SP $p$ receives requests at rate $\lambda_{c,p}=\lambda\cdot f_p\cdot \zeta_p\cdot \rho_{c,p}$. As usually done in the literature, we assume all objects have the same size. They may represent, for instance, chunks of videos.
We assume that the sequence of requests is a stationary stochastic process.

It is reasonable to assume that object popularity and request rate change smoothly over time, and we verify that our algorithm converges in a small lapse of time (1h), during which we assume popularity to be stationary.

\subsection{Cache Partitioning}
\label{sec:Cache-partition}

\begin{figure}[t]
    \centering
    \includegraphics[width=0.35\textwidth]{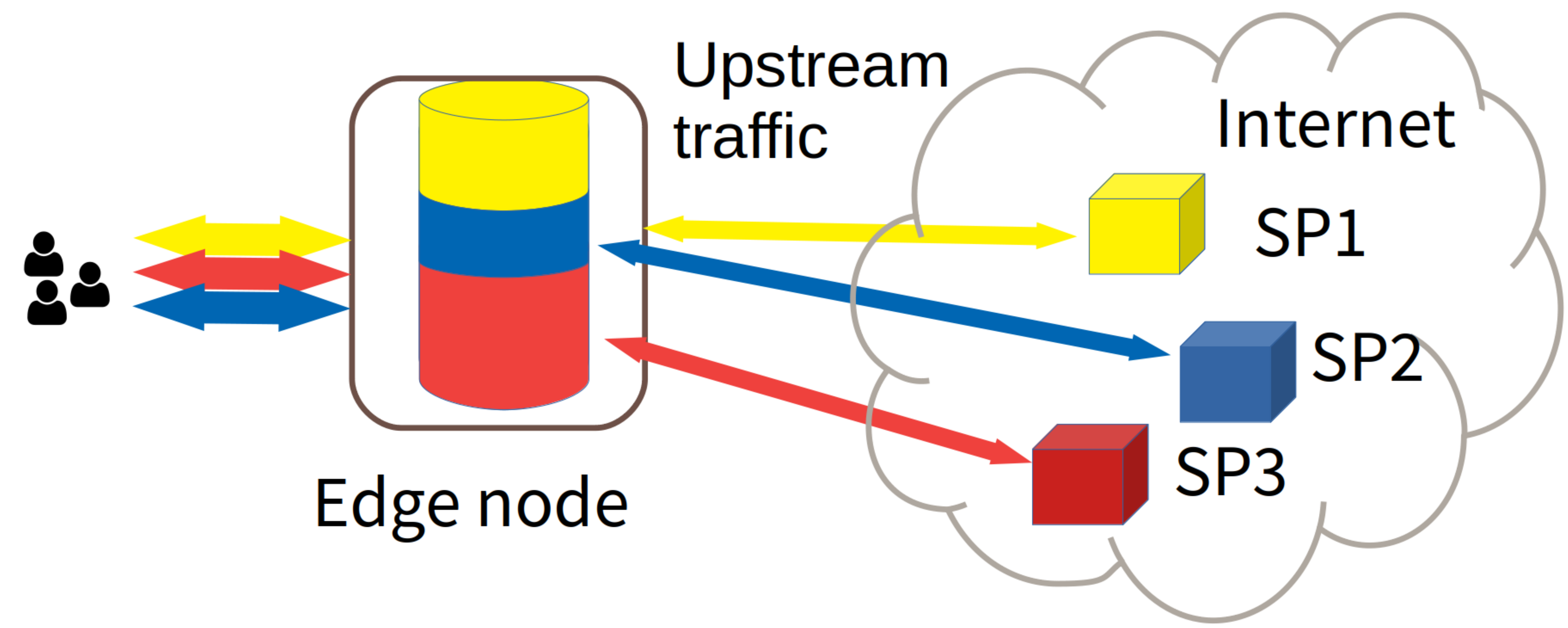}
    \caption{Cache allocation and upstream traffic (Origin servers $\rightarrow$ Edge) with multiple Service Providers (SPs).}
    \label{fig:upstream-traffic}
\end{figure}

The NO owns storage resources at the edge of the network, e.g., in a server co-located with a (micro) base station, and aims to minimize the inter-domain upstream traffic arriving from other Autonomous Systems (ASes), which it generally pays for. To do so, it allocates a total storage of $K$ slots among the $P$ SPs. For the sake of clarity, we focus on a case where each SP is a video streaming service, which caches its most popular objects (videos) in its allocated slots, but our results could be generalized to other situations. We assume one slot can store one object. The allocation is a vector $\boldsymbol{\theta}$ = ($\theta_1$,...,$\theta_P$) where each $\theta_p$ is the number of slots given to SP $p$ and $\sum_{p=1}^{P} \theta_p \leq K$. We define the set of all possible allocations as:
\begin{align}
    \label{eq:T}
    \mathcal{T}\triangleq
    \left\{
    \boldsymbol{\theta} | \sum_{p=1}^P \theta_p\le K, \theta_p\in\mathbb{Z}^+
    \right\}.
\end{align}

Whenever $\theta_p$ slots are given to SP $p$, it caches there its $\theta_p$ most popular objects.
The time is slotted and at any time slot the NO may perturb the allocation, giving $\Delta$ slots to one SP and subtracting $\Delta$ slots from another. 


\subsection{Cost Model}
\label{sec:cost-Model}
A user request for an object contained into the edge cache is served directly by the edge, otherwise the object must arrive from another AS, generating inter-domain upstream traffic. We call the amount of requested objects that must be retrieved upstream the \emph{nominal cost}. The nominal cost is a stochastic quantity that depends not only on the decided allocation $\boldsymbol{\theta}$, but also on exogenous conditions, which the NO cannot control, e.g., the amount of requests of the users of each SP. We denote exogenous conditions with a random variable $\omega$.
We denote the nominal cost as $C_{\text{nom}}(\boldsymbol{\theta},\omega)$. 

We assume that SPs are isolated and thus this cost can be decomposed into a sum of costs $C_{\text{nom},p}(\boldsymbol{\theta},\omega)$, each one quantifying the upstream traffic of one SP:
\begin{align}
\label{eq:decomposition}
    C_{\text{nom}}(\boldsymbol{\theta},\omega) = 
    \sum_{p=1}^P C_{\text{nom},p}(\boldsymbol{\theta},\omega).
\end{align}

Thus, we assume that the cost generated by SP $p$ only depends on $\theta_p$, i.e., $C_{\text{nom},p}(\boldsymbol{\theta},\omega) = C_{\text{nom},p}(\theta_p,\omega)$.

The NO can monitor the amount of traffic of a certain SP from the edge node to the users. A part (i) of this traffic will originate from the SP cache located at the edge node and the other part (ii) from the remote servers of the SP somewhere in the Internet (Fig.~\ref{fig:upstream-traffic}). The traffic saved corresponds to the difference between (ii) and (i).

\subsection{Data-Driven Optimization}
\label{sec:Optimization-Model}

The NO wants to solve the following \emph{optimal allocation problem}:
$
\boldsymbol{\theta}^* \in \arg\min_{\boldsymbol{\theta}\in\mathcal{T}}  C_{\text{nom}}(\boldsymbol{\theta},\omega)
$.
If the expression of the cost $C_{\text{nom},p}(\cdot,\cdot)$ for any SP $p$ and the exogenous conditions $\omega$ were known in advance by the NO, one could have aimed at solving such an optimization problem in an exact way by means, for instance, of dynamic programming. However, such information is not known, which renders the optimization problem above challenging to solve. Observe that for any $\boldsymbol{\theta}$, the total cost $C_{\text{nom}}(\boldsymbol{\theta},\omega)$ is a random variable depending on the exogenous random parameter $\omega$. Therefore, the NO can at most try to minimize its expected value:
\begin{align}
\label{eq:optim}
    \boldsymbol{\theta}^* 
    \in \arg\min_{\boldsymbol{\theta} \in \mathcal{T} }  \mathbb{E}C_{\text{nom}}(\boldsymbol{\theta})
\end{align}

However, not even this last optimization problem is directly solvable, since the NO does not know the form of the function $\boldsymbol{\theta}\rightarrow \mathbb{E}C_{\text{nom}}(\boldsymbol{\theta})$. This is why we resort to data-driven approach: at each time-slot, we perturb the allocation $\boldsymbol{\theta}$ by vector $\mathbf{a}$, which we call \emph{perturbation vector} and which denotes the amount of cache slots we add or remove to each SP (i.e., $\boldsymbol{\theta}:=\boldsymbol{\theta}+\mathbf{a}$). Then, we measure the effect of the perturbation, i.e., we measure the new $C_{\text{nom}}(\boldsymbol{\theta},\omega)$. We emphasize that $\mathbb{E}C_{\text{nom}}(\boldsymbol{\theta})$ that we would aim to minimize in~\eqref{eq:optim} is never observable directly, only $C_{\text{nom}}(\boldsymbol{\theta}, \omega)$ is. The latter can be considered as a noisy observation of $\mathbb{E}C_{\text{nom}}(\boldsymbol{\theta})$, where the noise is:
\begin{align}
\label{eq:noise}
    n=\mathbb{E}C_{\text{nom}}(\boldsymbol{\theta})-C_{\text{nom}}(\boldsymbol{\theta},\omega)
\end{align}

Every perturbation $\mathbf{a}$ produces spurious upstream traffic. Indeed, if SP $p$ had $\theta_p$ slots in the previous time-slot and it has $\theta_p+\Delta$ cache slots after a perturbation, it has to download the $\Delta$ new objects to fill the newly granted cache slots. This generates an upstream traffic corresponding to $\Delta$ objects. We call this traffic \emph{perturbation cost} and denote it by $C_{\text{pert}}(\mathbf{a})$. Note that this quantity is deterministic. For any time slot $k$, let us denote with $\boldsymbol{\theta}^{(k)}, \mathbf{a}^{(k)}, \omega^{(k)}$ the current allocation vector, the perturbation applied by the NO and the realization of the exogenous conditions, respectively. The cumulative cost over $Z$ time-slots is thus the sum of \emph{instantaneous costs} $C^{(k)}$, defined as follows:
\begin{align}
\label{eq:C_cum}
    C_{\text{cum}}(Z) &
    =  \sum_{k=1}^Z
    C^{(k)}
    \\
        \label{eq:instantaneous-cost}
    \text{where instantaneous cost }
    C^{(k)}
    & \defeq 
    C_{\text{nom}}(\boldsymbol{\theta}^{(k)},\omega^{(k)})
        + C_{\text{pert}}(\mathbf{a}^{(k)})
\end{align}

Note that, despite the fact that the spurious traffic generated by perturbations adds to the cost, perturbations are the only way for the NO to discover how to optimize the ``black-box'' function $\boldsymbol{\theta}\rightarrow \mathbb{E}C(\boldsymbol{\theta})$. Indeed, by observing the effects of perturbations on the nominal cost, the NO can accumulate knowledge that it can use to drive the system close to the optimal allocation $\boldsymbol{\theta}^*$.
Therefore, in our data-driven approach, rather than directly solving~\eqref{eq:optim}, which would be infeasible for the reasons stated above, our aim is to find a sequence of perturbations $\mathbf{a}^{(k)}$ in order to minimize~\eqref{eq:C_cum}.
Thus, we resort to RL (detailed in~\S~\ref{sec:rl-formulation}). In the numerical results, we show that, by doing so, we nevertheless approach the optimal allocation~\eqref{eq:optim}. Note that, for any initial allocation $\boldsymbol{\theta}^{(0)}$, the sequence $\{\mathbf{a}^{(k)}\}$ deterministically induces a sequence of states $\{\boldsymbol{\theta}^{(k)}\}$:
\begin{align}
\label{eq:dynamics}
    \boldsymbol{\theta}^{(k+1)}=\boldsymbol{\theta}^{(k)}+\mathbf{a}^{(k+1)}.
\end{align}

\section{Reinforcement Learning Formulation}
\label{sec:rl-formulation}

\subsection{General setting}
\label{sec:General-setting}
We make use of RL to solve the data-driven cache allocation problem described above. The set of \textbf{states} $\mathcal{S}$ consists of all the allocation vectors that we can visit. To reduce the complexity of the problem, we adopt a discretization step $\Delta\in\mathbb{N}$, and define $\mathcal{S}$ as:
\begin{align}
    \label{eq:state}
    \mathcal{S}=
    \left\lbrace \pmb{\theta} = (\theta_1,\dots,\theta_P) |  \sum_{p=1}^P \theta_p \le K, \theta_p \text{ multiple of }\Delta
    \right\rbrace
\end{align}

The discretization step $\Delta$ constitutes a precision/complexity trade-off. A smaller value of $\Delta$ increases the precision of the allocation since it allows to converge to a discrete solution closer to the optimal one (\S~\ref{sec:pre-tuning-hyper-parameters}); it however increases the complexity of the problem since it expands the space of states. 

Observe that $\mathcal{S}\subset\mathcal{T}$ (see~\eqref{eq:T}).
When in state $\pmb{\theta}$, the NO can pick an action from the following \textbf{action space}:
\begin{align}
\label{eq:action-space}
    \mathcal{A}_{\pmb{\theta}} = \left\lbrace
        \mathbf{a} = 
        \Delta\cdot (\mathbf{e}_p - \mathbf{e}_{p'} ) |
        \pmb{\theta} + \mathbf{a}\in\mathcal{S},
        p,p'=1,\dots,P
    \right\rbrace
\end{align}
where $\mathbf{e}_p$ is the $p$-th element of the standard basis of $\mathbb{R}^p$.

We will use the terms allocation/state and action/perturbation interchangeably.
Therefore, an action $\mathbf{a}$ consists in the NO adding 
$\Delta$
units of storage to a certain SP $p$ and removing the same amount from another SP. The null action corresponds to not changing the allocation (which happens in~\eqref{eq:action-space}  when $p=p'$). Thanks to~\eqref{eq:dynamics}, the transition from a state to another is deterministic.

Our objective function accounts for both nominal cost as well as perturbation cost and is given by: 
 
 \begin{scriptsize}
  \begin{align}
\label{eq:C_cum_disc}
    C_{\text{cum}}^\gamma =
    \lim_{Z\rightarrow \infty}
    \mathbb{E}
    \left[
    \sum_{k=0}^Z \gamma^{(k)} \cdot
    \underbrace{\left(
        C_{\text{nom}}(\boldsymbol{\theta}^{(k)},\omega^{(k)})
        + C_{\text{pert}}(\mathbf{a}^{(k)})
    \right)}_{\text{Instantaneous cost } C^{(k)}}
    \right]_{
        \begin{array}{c}
        \boldsymbol{\theta}^{(k)}\in\mathcal{S}\\
        \mathbf{a}^{(k)}\in\mathcal{A}
        \end{array}
    }
\end{align}
 \end{scriptsize}
where $\gamma<1$ is a hyper-parameter called \emph{discount factor}.

A policy $\pi$ is a function $\pi(\mathbf{a}|\boldsymbol{\theta})$ defining the decisions of the NO: whenever the NO observes state $\boldsymbol{\theta}$, it will choose an action $\mathbf{a}$ with probability $\pi(\mathbf{a}|\boldsymbol{\theta})$. During training, the NO starts with a certain policy $\pi^{(0)}(\cdot)$ and then adjusts it, based on measured cost, in order to approach the optimal policy $\pi^*(\cdot)$, i.e., the one that minimizes~\eqref{eq:C_cum_disc} (\S~7 of~\cite{Tsitsiklis1994}). Therefore, at any iteration $k$, function $\pi^{(k)}(\cdot)$ evolves. In particular, at any time slot $k$ the NO observes the current state $
\boldsymbol{\theta}^{(k)}$, takes an action $\mathbf{a}^{(k)}$ probabilistically, according to the current policy $\pi^{(k)}(\mathbf{a}|\boldsymbol{\theta}^{(k)}), \forall \mathbf{a}\in\mathcal{A}_{\boldsymbol{\theta}^{(k)}}$. Then, the instantaneous cost $C^{(k)}$ is measured. Such a measurement is adopted to improve policy $\pi^{(k)}(\mathbf{a}|\boldsymbol{\theta})$, which thus becomes $\pi^{(k+1)}(\mathbf{a}|\boldsymbol{\theta})$. The next section explains how such an improvement is obtained.

\subsection{Q-Learning}
\label{sec:q-learning}
 Among the different flavors of RL, we chose Q-learning, which has the advantage of being easy to implement and adapt to different problems (\S~4.3.1 of~\cite{Szepesvari2010}). A \emph{Q-table} is maintained, which associates to any pair $(\boldsymbol{\theta},\mathbf{a})$ a value $Q(\boldsymbol{\theta},\mathbf{a})$ that approximates the cumulative cost~\eqref{eq:C_cum_disc} when being at the state $\boldsymbol{\theta}$ and choosing the action $\mathbf{a}$. This approximation is continuously improved based on the observed instantaneous cost $C^{(k)}$. In particular, at every time-slot $k$, the Q-table is updated as follows:

\begin{scriptsize}
\begin{flalign}
\nonumber
    Q(\boldsymbol{\theta}^{(k)},\mathbf{a}^{(k)}) 
    \\ 
    := (1-\alpha^{(k)})\cdot
    Q(\boldsymbol{\theta}^{(k)},\mathbf{a}^{(k)})  
    + \alpha^{(k)} \cdot \left( C^{(k)} + \gamma \min_{\mathbf{a}\in\mathcal{A}_{\boldsymbol{\theta}^{(k+1)}}} Q(\boldsymbol{\theta}^{(k+1)},\mathbf{a}) 
    \right)
    \label{eq:q-learning}
\end{flalign}  
\end{scriptsize}

The Q-table entirely determines the policy, in the sense that at any time-slot $k$ we choose a random action $\mathbf{a}^{(k)}\in\mathcal{A}_{\boldsymbol{\theta}^{(k)}}$ with probability $\epsilon^{(k)}\in[0,1]$ and the ``best'' action $\mathbf{a}^{(k)}=\arg\min_{\mathbf{a}\in\mathcal{A}_{\boldsymbol{\theta}^{(k)}}}$ $Q(\boldsymbol{\theta}^{(k)},\mathbf{a})$ with probability $1-\epsilon^{(k)}$. This is the so-called $\epsilon$-greedy algorithm. 

\subsection{Additional Enhancement}
\label{sec:additional-enhancement}

We now report some enhancements to Q-learning that considerably improved the performance of our algorithm (\S \ref{sec:pre-tuning-hyper-parameters}):

\textbf{(I)} The parameter $\alpha^{(k)}$ in~\eqref{eq:q-learning} is \emph{learning rate}. As in \cite{Araldo2018}, we decrease it  slowly, to keep Q-table updates relatively large:
\begin{align}
    \alpha^{(k)} = \alpha^{(k-1)} \cdot 
    \left(1 - \frac{1}{1 + M + k}\right)^{\frac{1}{2} + \xi}
    \label{eq:alfa-function}
\end{align}
where $M$ and $\xi$ are positive constants, used to tune the slope of decrease. 

\textbf{(II)} In the simplest implementation of Q-learning, the measurement made in a certain time-slot is used to update the Q-table in that time-slot only and is never used again. However, the set of previous measurements (i.e., the past ``experience'') could be further exploited to improve the Q-table update in future time-slots. To this aim, Experience Replay has been proposed \cite{William2020}.
At any time-slot $k$, in addition to using the measured instantaneous cost $C^{(k)}$ to update the Q-table in~\eqref{eq:q-learning}, we also store this measurement in the form of a triplet $\left(\boldsymbol{\theta}^{(k)}, \mathbf{a}^{(k)}, C^{(k)}\right)$, which we call \emph{experience}. The set of experiences accumulated in this way is called \emph{memory}. Whenever we update the Q-table, additionally to performing~\eqref{eq:q-learning} using the current observation, we also sample the memory randomly for a mini-batch of experiences of size $N$ and we use them when applying~\eqref{eq:q-learning}.

\textbf{(III)} The value of $\epsilon^{(k)}$ is the probability of taking a random action, instead of the best so far, at any time-slot $k$. We impose, motivated by \cite{natarajan2020}, the following decay:

\begin{align}
    \epsilon^{(k)} =
    \begin{cases}
    \epsilon_0 - \left[  \frac{ 0.9 \cdot \epsilon_0 } { \cosh (e^{-\frac{k-A\cdot Z}{B\cdot Z}}) } + \frac{k \cdot C}{Z} \right]
    & \text{ if }k\le Z
    \\
    \frac{\epsilon^{(Z)}}{k-Z} & \text{ otherwise }
    \end{cases}
    \label{eq:epsilon-formula}
\end{align}
where $\epsilon_0$ is the initial value of $\epsilon$, $A$, $B$ and $C$ are hyper parameters and $Z$ is a time horizon.
This decay provides:
(i)~sufficient time for exploration at the beginning, 
(ii)~preference to exploitation (with respect to exploration) in the end (quasi-deterministic policy) and
(iii)~smooth transition while switching from exploration to exploitation. $A$ decides whether to spend more time on exploration or on exploitation, $B$ decides the slope of the transition between them and $C$ decides the steepness of the $\epsilon^{(k)}$ decay.

\subsection{Discussion on the use of RL}
\label{sec:methodology}
We now briefly discuss why we preferred our RL setting over other possible methodologies.
First of all, we rule out all static optimization techniques that require full information, due to the online and stochastic nature of the problem at hand.

We could also interpret our allocation problem as a ``black-box optimization'' and apply Bayesian Optimization~\cite{DeFreitas2016}. However, such techniques are meant for offline problems, where the objective is to retrieve the minimum of the cost function \emph{at the end of the optimization}~\eqref{eq:optim} and the cost of jumping from one state to another is neither quantified nor directly minimized. Our RL framework not only allows us to reach an allocation close to the optimum at the end, but also implicitly optimizes the path of states visited \emph{during} the optimization.

Lyapunov Optimization (LO) has also been used for allocation problems~\cite{Giannakis2017} but it assumes some knowledge about the expression of the stochastic reward function. We instead optimize the system even if it is unknown.

The Markov Decision Process (MDP) underlying our RL method is a Deterministic MDP (DMDP), as the transition from one state to another is deterministic~\eqref{eq:dynamics}. In~\cite{Lazaric2018}, DMDP is solved assuming the structure of the reward function is known, which in our case we do not know. 

If we wanted to apply Multi-Armed Bandit (MAB), we would need to interpret each allocation vector as an arm.  However, MAB does not allow to consider the cost of ``jumping'' from one arm to another.  

Online decision problems have been presented in an adversarial setting and solved via Smoothed Online Convex Optimization (SOCO)~\cite{Goel2019}.
In adversarial setting, performance bounds are calculated in a worst-case analysis. In such a setting, \cite[Theor.3.1]{Dekel2013} shows that it is impossible to effectively optimize a DMDP such as ours. We instead adopt a stochastic setting and study the ``average'' behavior of the system.

\section{Numerical Results}
\label{sec:performance-evaluation}
We now evaluate the performance of our RL allocation $\boldsymbol{\theta}^{(k)}$ through simulations developed in Python and compare it with two static allocations: (i) the theoretical optimal allocation $\boldsymbol{\theta}^{*}$, which would ideally be computed by an oracle who knows exactly the content popularity and thus the expression of function $\boldsymbol{\theta}\rightarrow \mathbb{E}C(\boldsymbol{\theta})$ and (ii) the proportional allocation $\boldsymbol{\theta}^{\text{prop}}$ where $\theta_p$ is proportional to the rate of requests $\lambda_p$ directed to SP $p$. We also compare our RL algorithm to SPSA~\cite{Araldo2018}.

We consider a network with 3 SPs. We set the overall request arrival rate to $\lambda = 4 \cdot 10^3 req/s$ (in the same order of magnitude of requests supported in one edge location of Amazon CloudFront). Each of these requests is directed to
SP $1$, $2$ or $3$ with probability $0.75, 0.20, 0.05$, respectively.
We set the cacheability (\S~\ref{sec:Request-patterm}) of $SP_1$, $SP_2$ and $SP_3$ to $\zeta_1=0.4,\zeta_2=0.9,\zeta_3=0.9$. Each SP has catalog of $N_1=N_2=N_3=10^7$ cacheable objects. Content popularity in each catalog follows Zipf’s law with exponent $\beta_1$ = 1.2, $\beta_2$ = 0.4 and $\beta_3$ = 0.2, respectively. The total cache size is $K = 5\cdot 10^6$.
The simulation time is set to 6 hours. The length of a time-slot is 0.25 second.

We plot a normalized cost, i.e., the amount of objects downloaded from the Internet (either as a result of an edge cache miss or of an allocation perturbation) divided by the total amount of objects requested by the users.
All curves are averaged with a sliding window of 10 min.

\subsection{Pre-tuning of hyper-parameters and convergence}
\label{sec:pre-tuning-hyper-parameters}
We now discuss some preliminary tuning that we performed, including the features indicated in \S\ref{sec:additional-enhancement}.

(1) For the discretization step $\Delta$, we found out that a good complexity vs. precision trade-off was to set it to $K/50$.
To limit perturbations, we give a higher ``weight'' to the null action. Indeed, when we take a random action (\S~\ref{sec:q-learning}), we set the probability of choosing any non-null action to only $1/P^2$ and all the remaining probability is for the null-action.

(2) For $\gamma$, we set it to 0.99, i.e., very close to 1 to give importance to future rewards and prevent myopic decisions. 

(3) For $\alpha$, the learning rate, we found that convergence was slow when it was fixed. Therefore, we adopt learning rate scheduling, which starts at 0.9 and decreases following (\ref{eq:alfa-function}) to 0.2, with $M=3600$ and $\xi=0.01$).

(4) Regarding the size $N$ of mini-batch of experiences, we found that small fixed values were not allowing to exploit past experience, on the other hand, with large values past experience was dominating too much the updates. We obtained the best performance by scheduling $N$ as follows:

\begin{align}
    {N}^{(k)} =
     \frac{N_{\text{max}}} { \cosh (e^{-\frac{k-A\cdot Z}{B\cdot Z}}) } + \frac{k \cdot C}{Z}
    \label{eq:mini-batch-formula}
\end{align}
where $N_{\text{max}}$ = 100, $A$ = 0.15, $B$ = 0.3, $C$ = 0.7, $Z$ = 6 hours.

(5) Finally, we make $\epsilon$ decay as in~(\ref{eq:epsilon-formula}) with $A = 0.3$, $B = 0.1$ and $C = 0.01$. These hyper-parameters have been chosen empirically after preliminary experimentation and provide a good compromise between exploration and exploitation.

\subsection{Convergence close toward the optimum}
\label{sec:convergence}

\begin{figure}[t]
\begin{subfigure}{.25\textwidth}
  \centering
  \includegraphics[width=1\linewidth]{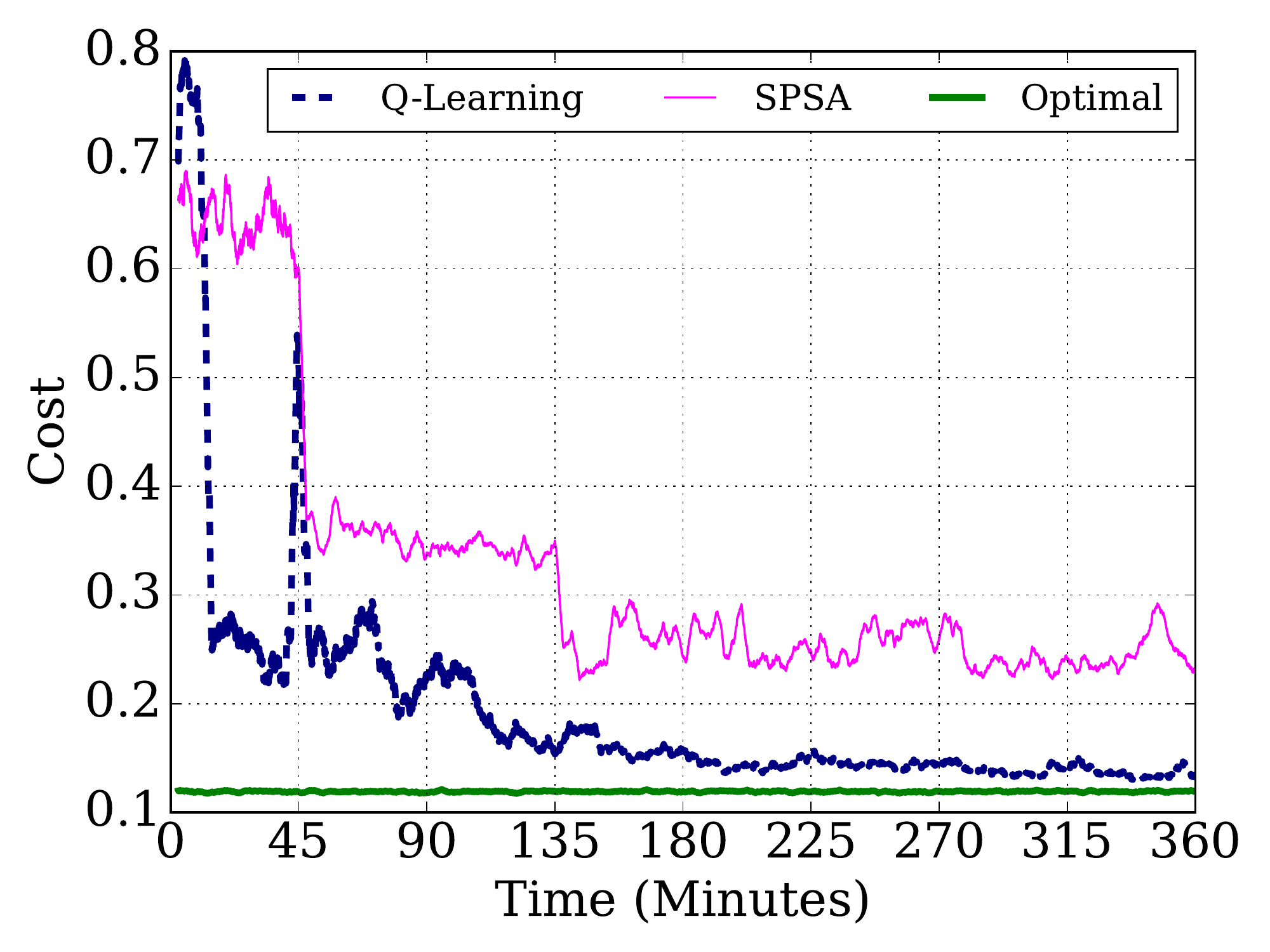}
  \caption{Total System Cost $C^{(k)}$}
  \label{fig:final-result-cost}
\end{subfigure}%
\begin{subfigure}{.25\textwidth}
  \centering
  \includegraphics[width=1\linewidth]{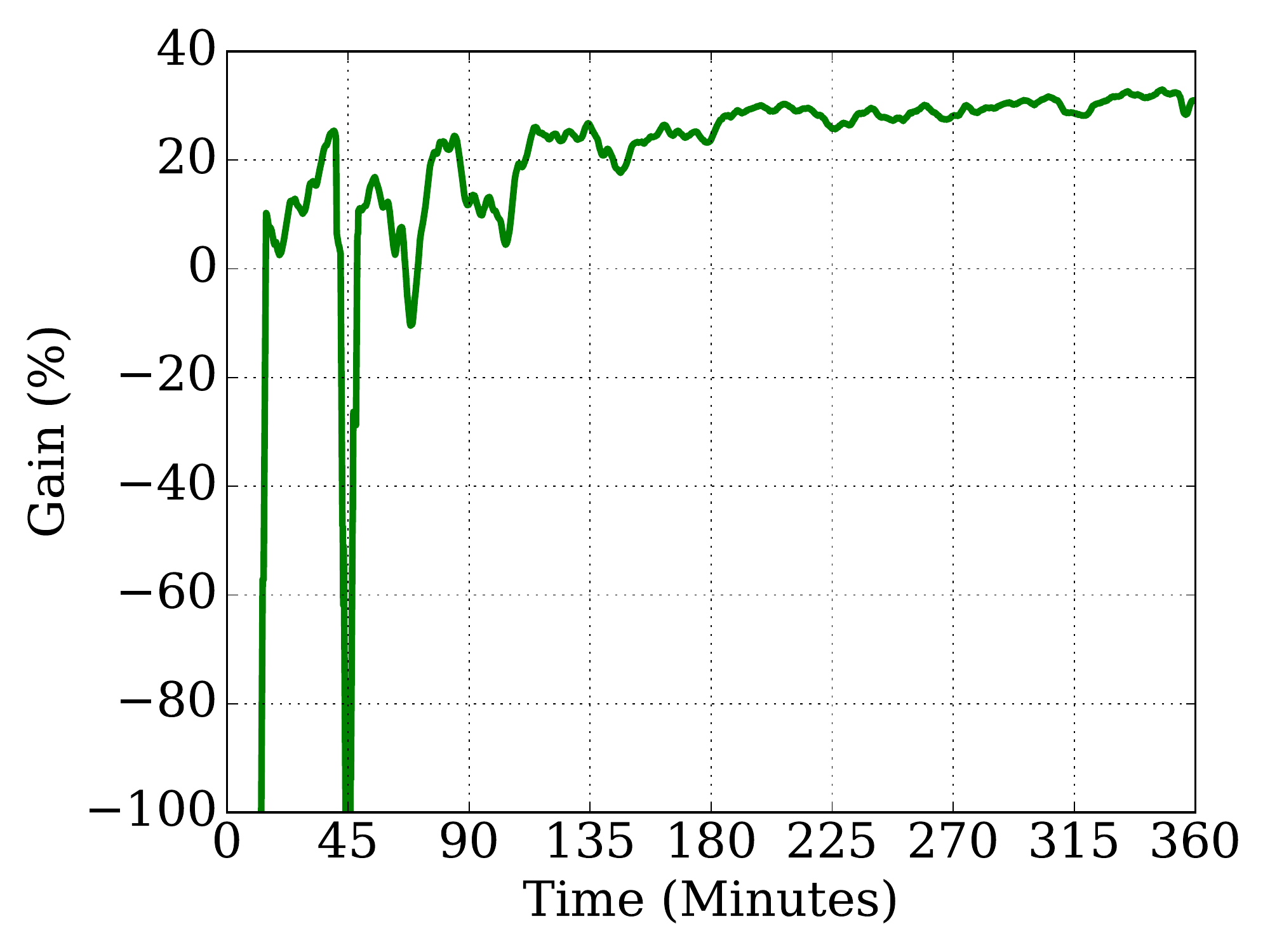}
  \caption{Gain with respect to $\boldsymbol\theta^{\text{prop}}$}
  \label{fig:gain-prop}
\end{subfigure}
\caption{System Performance}
\label{fig:system-3SP}
\end{figure}

The behavior of our algorithm is well illustrated by Fig.~\ref{fig:system-3SP}: in a first phase, we know nothing about the system and we need to perturb it a lot by taking many random actions, in order to learn it. For this reason, perturbation cost is high up to 135 minutes. After that, we start to exploit the collected knowledge and we limit perturbation. This ``explore-then-exploit'' behavior is very effective in rapidly reducing overall cost (Fig.~\ref{fig:final-result-cost}) toward the theoretical optimum. 

Furthermore, our RL algorithm outperforms SPSA used in~\cite{Araldo2018} which converges to the optimal allocation in 45 minutes but never reaches the optimum due to the continuous perturbations it has to apply to estimate the sub-gradient of the objective function.   

We now compare the cost collected by our policy $C^{(k)}$ with the cost of the static allocation $\boldsymbol{\theta}^{\text{prop}}$. Note that while our method deals with both nominal and perturbation costs~\eqref{eq:instantaneous-cost}, the static $\boldsymbol{\theta}^{\text{prop}}$ does not apply any perturbation to the system. We define the gain of our policy with respect to $\boldsymbol{\theta}^{\text{prop}}$ as:
\begin{small}
\begin{align}
\label{eq:gain}
    G_{\text{prop}}^{(k)}=
    \frac{ 
        C_{\text{nom}}(\boldsymbol{\theta}_{\text{prop}},\omega^{(k)}) 
        - C^{(k)}
    }
    {
        C_{\text{nom}}(\boldsymbol{\theta}_{\text{prop}},\omega^{(k) }
        )
    }
\end{align}
\end{small}

Fig.~\ref{fig:gain-prop} shows that our solution reaches a gain of 29\% in less than 3 hours with respect to $\boldsymbol{\theta}^{\text{prop}}$.

\subsection{Fairness}
\label{sec:fairness}

Let us denote with $x_p = \frac{\theta_p}{\zeta_p \cdot \lambda \cdot f_p}$ the slots given to SP $p$, normalized to its amount of cacheable requests.
We compute the fairness of the system with the Jain's fairness index as
$\mathcal{J}(x_1, \dots, x_P)\triangleq
\frac{(\sum_{p=1}^P x_p)^2}{P\cdot\sum_{p=1}^P x_{p}^2}$.

Our results show that cache sharing strategy with our RL-based allocation $\boldsymbol{\theta}^{\text{RL}}$ (0.7 fairness) is much fairer than the optimal allocation $\boldsymbol{\theta}^*$ (0.36 fairness), at almost the same total cost. It is also close to that of the proportional allocation $\boldsymbol{\theta}^{\text{prop}}$ (0.85 fairness) albeit being much better in terms of cost. 
Note that we are also close to the ideal maximum fairness achieved by the proportional allocation not taking into account cacheability, i.e. if all contents were cacheable (i.e. $\zeta_p = 1, p = 1,..,P$).  The latter is 1, by construction, as it is proportional to the rate of requests directed to each SP; on the other hand, it is an artificial measure, as it ignores cacheability.

\subsection{Sensitivity Analysis}
\label{sec:sensitivity-analysis}
\begin{figure}[t]
\begin{subfigure}{.25\textwidth}
  \centering
  \includegraphics[width=1\linewidth]{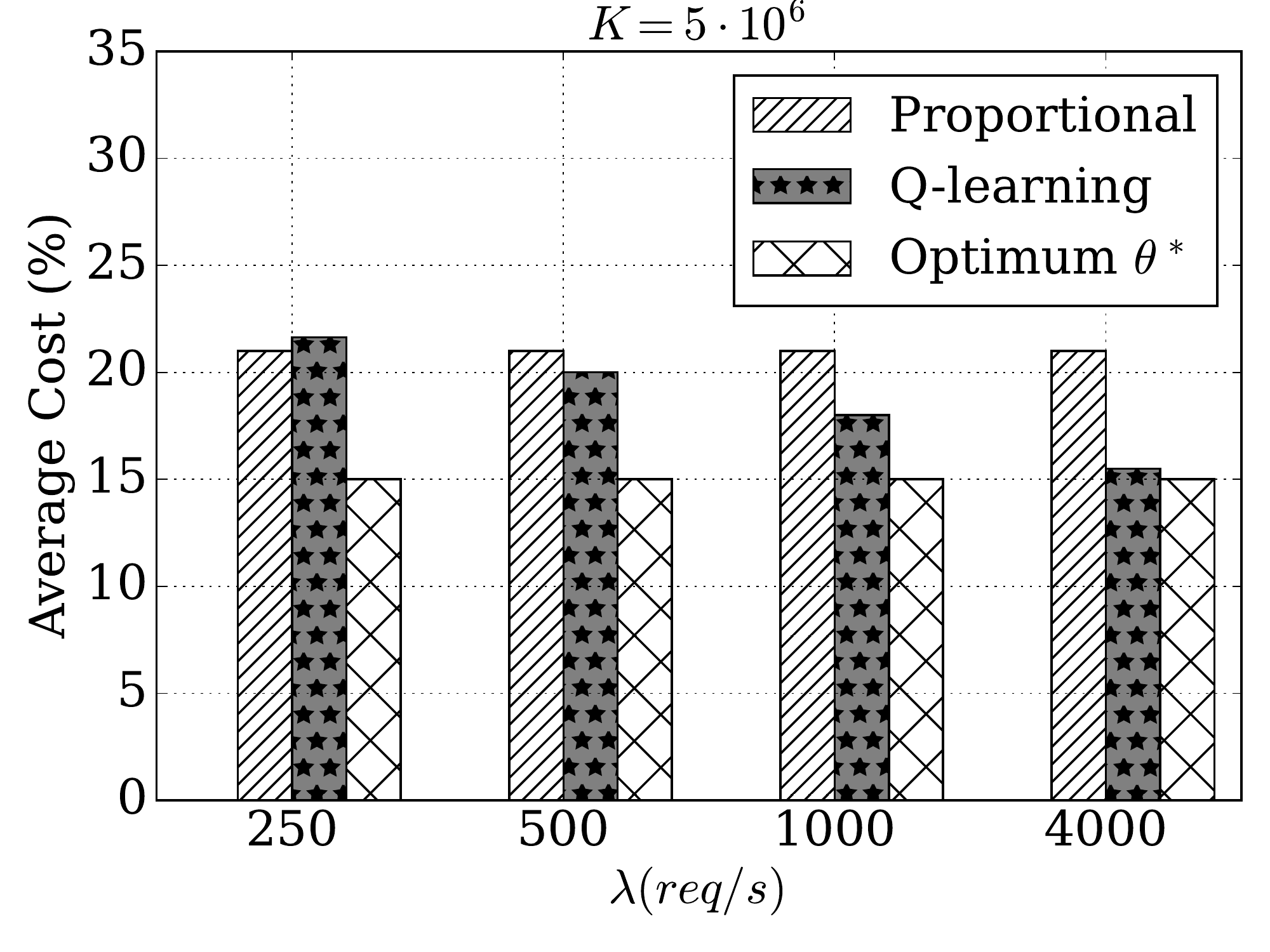}
  \caption{To request rate $\lambda$}
  \label{fig:sens-lambda}
\end{subfigure}%
\begin{subfigure}{.25\textwidth}
  \centering
  \includegraphics[width=1\linewidth]{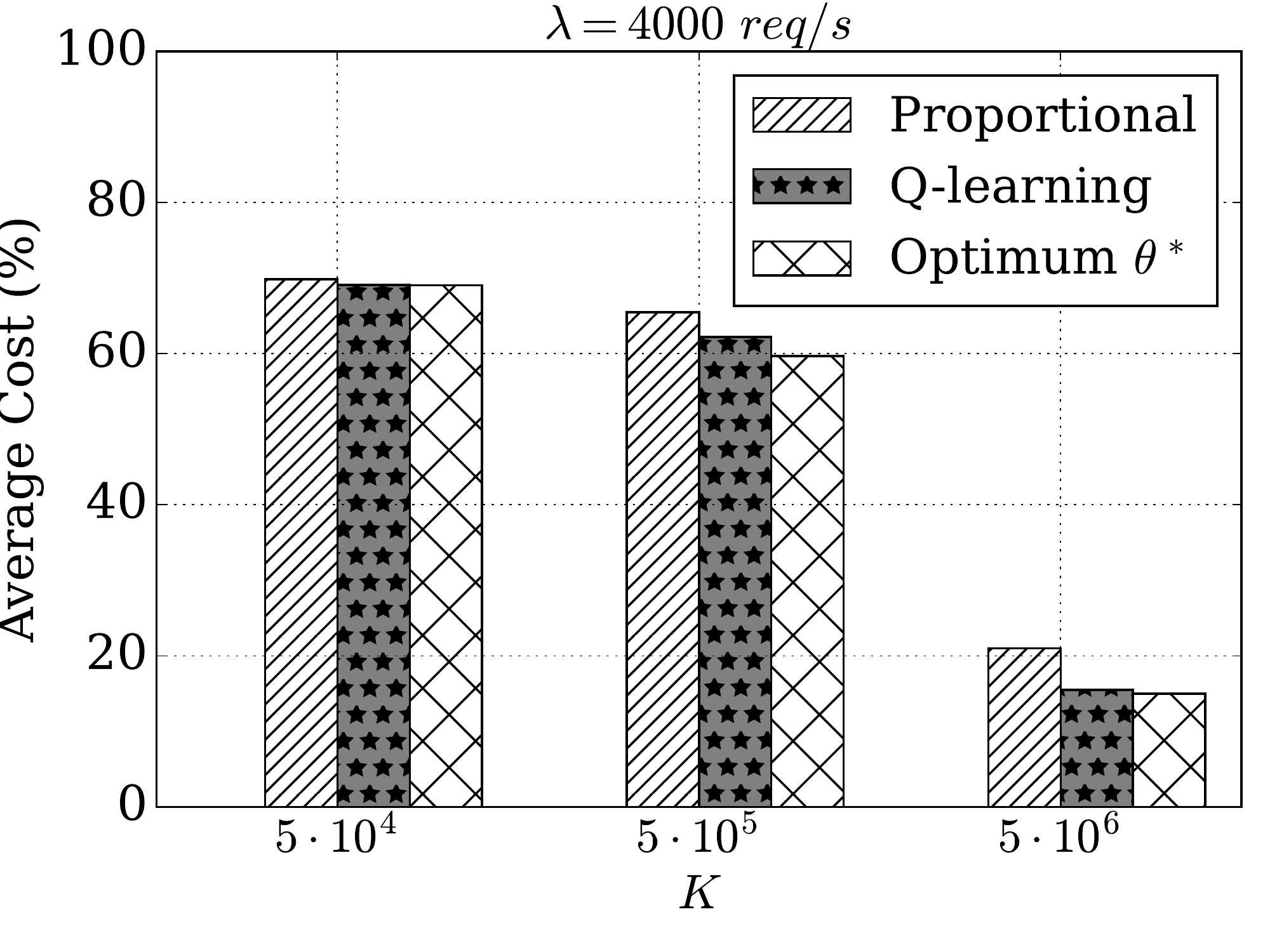}
  \caption{To cache size $K$}
  \label{fig:sens-K}
\end{subfigure}
\caption{Sensitivity of the system}
\label{fig:sensitivity}
\end{figure}

We next study how the performance of our solution is affected by the algorithm parameters and the scenario. We first focus on the request rate $\lambda$. Indeed, a small $\lambda$ implies that only few requests are observed in each time slot, which may result in a high noise, as defined in~\eqref{eq:noise}, and ultimately affects the accuracy of the update and slows down the convergence. We thus expect our Q-learning approach and, more generally, any data-driven approach, to perform best only with large $\lambda$.
This is confirmed by Fig. \ref{fig:sens-lambda}, where we plot the average cost $\frac{1}{Z} C_{\text{cum}}(Z)$~\eqref{eq:C_cum} of our RL algorithm, after $Z=6$ hours, and compare it to the static proportional and optimal allocations.

Fig. \ref{fig:sens-K} shows the average cost measured over $Z=6$ hours for various cache sizes $K \in \{5\cdot 10^4, 5\cdot 10^5, 5\cdot 10^6\}$ and a fixed request rate $\lambda = 4 \cdot 10^3 req/s$. It confirms that the gains of our algorithm hold for different cache sizes, and shows that gain increases for larger caches. Indeed, for small cache size there is not much to optimize: the cost is high with both proportional and optimal allocation, so even if our algorithm positions itself between the two, the cost saved is negligible.

\begin{figure}
\begin{subfigure}{.25\textwidth}
  \centering
  \includegraphics[width=1\linewidth]{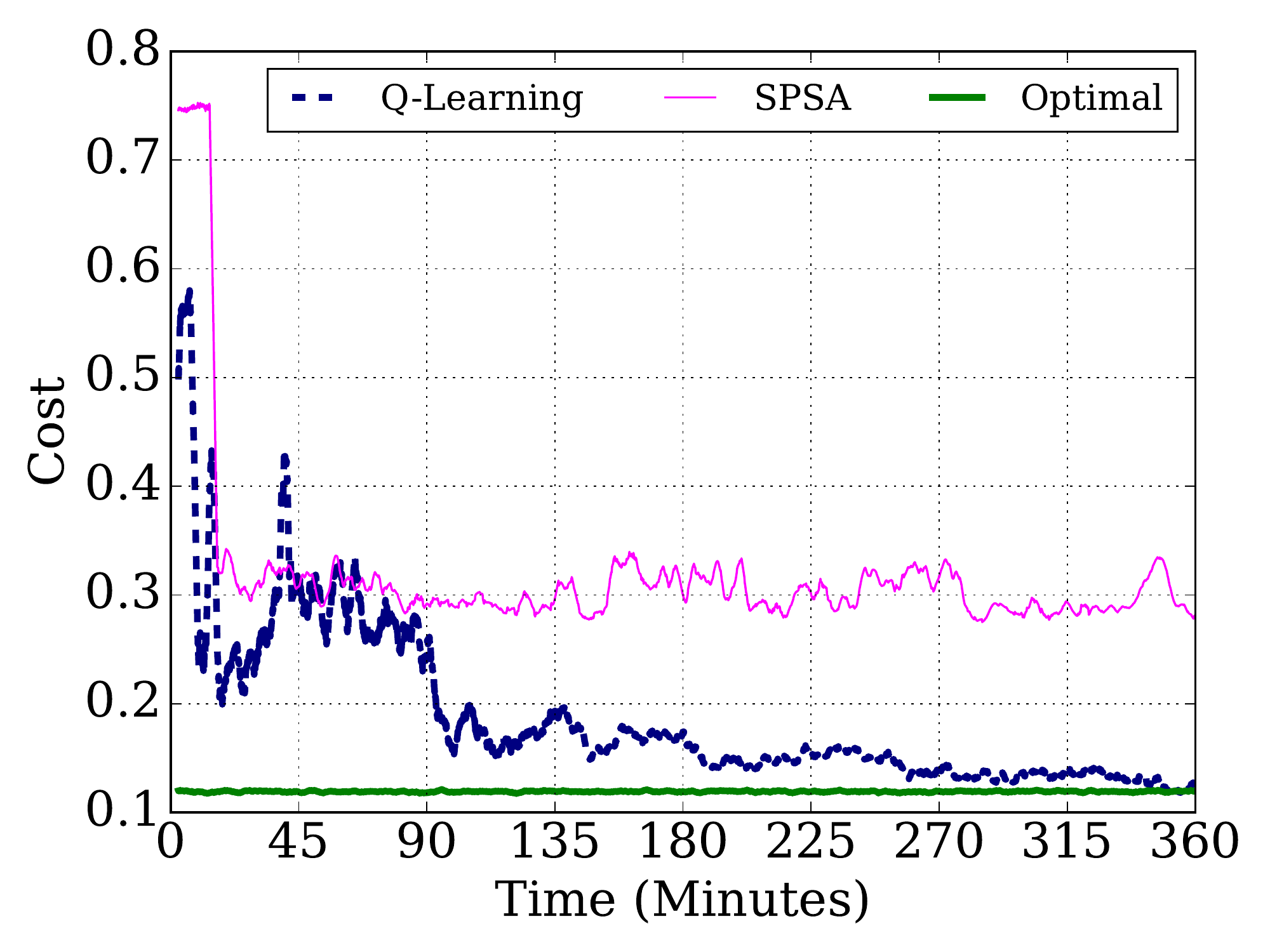}
  \caption{Total System Cost $C^{(k)}$}
  \label{fig:final-result-cost-4SP}
\end{subfigure}%
\begin{subfigure}{.25\textwidth}
  \centering
  \includegraphics[width=1\linewidth]{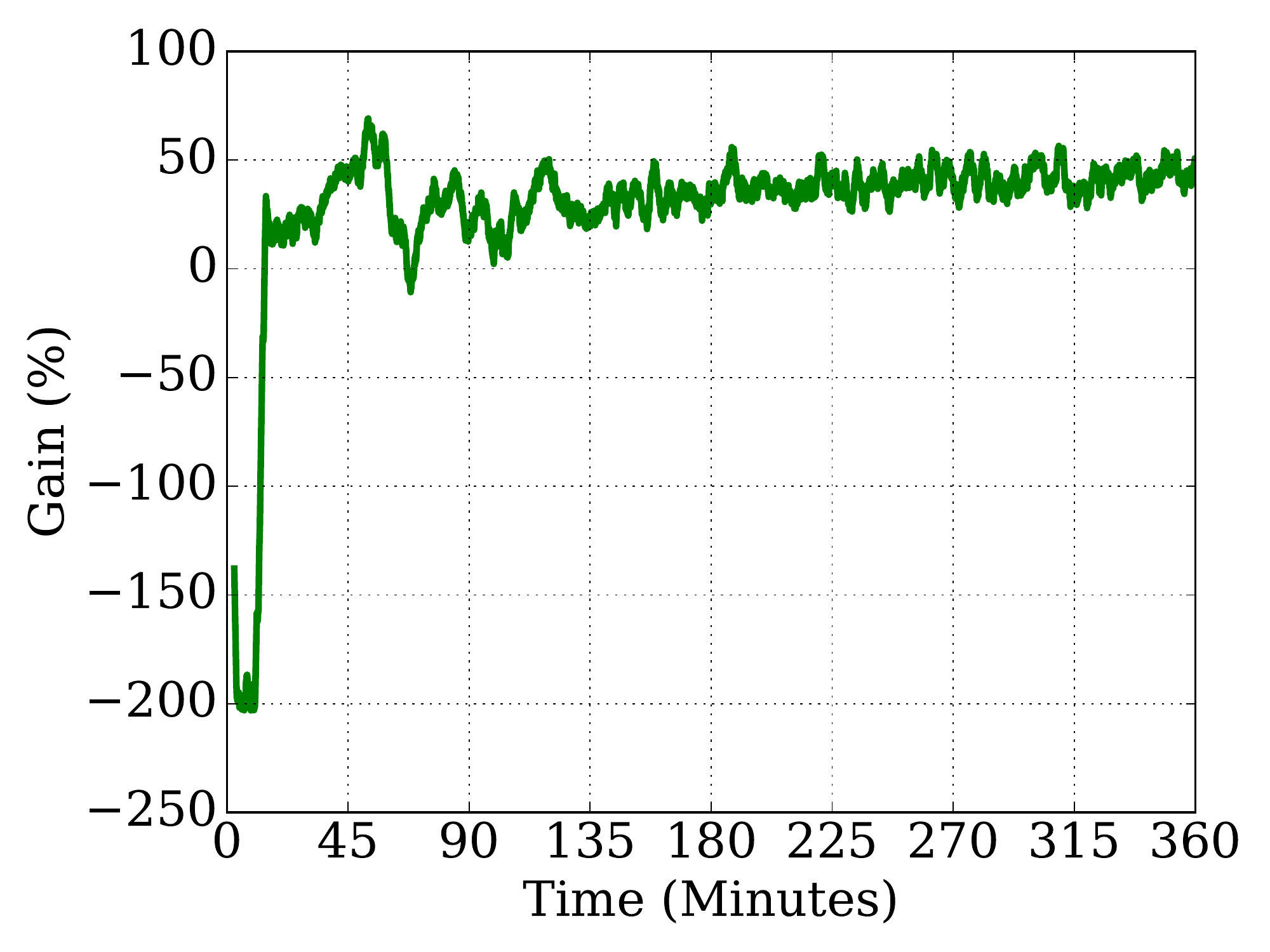}
  \caption{Gain with respect to $\boldsymbol\theta^{\text{prop}}$}
  \label{fig:gain-prop-4SP}
\end{subfigure}
\label{system-4SP}
\caption{System Performance for 4 SPs}
\end{figure}

We now study how our algorithm is affected by the number of SPs. We simulate the same scenario in the same conditions as in \S~\ref{sec:convergence} but we change the number of SPs to $P = 4$. As in \S~\ref{sec:convergence}, we plot in Fig.~\ref{fig:final-result-cost-4SP} the total cost $C^{(k)}$ of our RL algorithm vs. the optimal solution $\boldsymbol\theta^*$. The results show that our algorithm rapidly converges close to optimal cost, outperforming SPSA as well.

In Fig.~\ref{fig:gain-prop-4SP}, we plot the gain defined by \eqref{eq:gain} for 4 SPs. Results show that our RL algorithm continues to outperform $\boldsymbol\theta^{prop}$ by reaching a gain of 50$\%$ in 3 hours.

\section{Conclusion and Future work}
\label{sec:concl}
In this paper, we proposed a Q-learning based algorithm for cache allocation at the edge between several SPs with encrypted, not all cacheable content: a main challenge of in-network caching. We compared our dynamic allocation to the theoretical optimal and to a static allocation proportional to the probabilities of requesting content from each SP. We showed that our algorithm converges quite fast to a configuration close to the optimal and outperforms the proportional allocation in several system configurations as well as the state-of-the-art SPSA.  As part of the future work, we intend to consider scenarios with time varying popularity and to extend our work to multiple resources e.g., storage, CPU, RAM, etc. We would also consider a distributed scenario with multiple edge nodes.

\section{Acknowledgement}
\label{sec:ack}
This work was partially carried out in the Plateforme THD, a Fiber-To-The-Home platform of Telecom SudParis.



\bibliographystyle{IEEEtran}
\bibliography{main}
\end{document}